\documentclass[aps,prc,twocolumn,superscriptaddress,showpacs,floatfix,nofootinbib]{revtex4-1}
\usepackage{amsmath,graphicx,color}

\begin{document}

\title{Skewness of elliptic flow fluctuations}

\author{Giuliano Giacalone}
\affiliation{Institut de physique th\'eorique, Universit\'e Paris Saclay, CNRS,
CEA, F-91191 Gif-sur-Yvette, France} 
\author{Li Yan}
\affiliation{Institut de physique th\'eorique, Universit\'e Paris Saclay, CNRS,
CEA, F-91191 Gif-sur-Yvette, France} 
\author{Jacquelyn Noronha-Hostler}
\affiliation{Department of Physics, University of Houston, Houston TX 77204, USA}
\author{Jean-Yves Ollitrault}
\affiliation{Institut de physique th\'eorique, Universit\'e Paris Saclay, CNRS, CEA, F-91191 Gif-sur-Yvette, France} 
\date{\today}

\begin{abstract}
Using event-by-event hydrodynamic calculations, we find that the fluctuations of the elliptic flow ($v_2$) in the reaction plane have a negative skew. 
We compare the skewness of $v_2$ fluctuations to that of 
initial eccentricity fluctuations. 
We show that skewness is the main effect lifting the degeneracy 
between higher-order cumulants, with negative skew corresponding to 
the hierarchy $v_2\{4\}>v_2\{6\}$  observed in Pb+Pb 
collisions at the CERN Large Hadron Collider. 
We describe how the skewness can be measured experimentally and show that hydrodynamics naturally reproduces its magnitude and centrality dependence. 
\end{abstract}
\maketitle

\section{Introduction}

Elliptic flow, $v_2$, is one of the key observables of ultrarelativistic heavy-ion collisions at BNL Relativistic Heavy Ion Collider~\cite{Ackermann:2000tr} and
CERN Large Hadron Collider~\cite{Aamodt:2010pa}. 
Its large magnitude suggests that the strongly-coupled system formed in these collisions behaves collectively as a fluid~\cite{Luzum:2008cw}. 
However, quantitative comparison between hydrodynamic calculations and experimental data is hindered by the poor knowledge of the early collision dynamics and of the transport properties of the quark-gluon plasma~\cite{Heinz:2013th}. 
Therefore, it is essential to identify {\it qualitative\/} features predicted by hydrodynamics which can be tested against experimental data. 

A crucial step in our understanding of collective motion has been the recognition that $v_2$ fluctuates event to  event~\cite{Miller:2003kd,Alver:2006wh}. 
Elliptic flow fluctuations are quantitatively probed by the cumulants~\cite{Borghini:2001vi}, $v_2\{k\}$, with $k=2,4,6,8$~\cite{Abelev:2014mda,Aad:2014vba,Khachatryan:2015waa}.  
One typically observes $v_2\{2\}>v_2\{4\}$ and almost degenerate values for $v_2\{4\}$, $v_2\{6\}$, and $v_2\{8\}$, corresponding to Gaussian fluctuations of $v_2$~\cite{Voloshin:2007pc}. 
A fine splitting (at the percent level) between $v_2\{4\}$ and
$v_2\{6\}$ is, however, observed for most centralities~\cite{Aad:2014vba}. 
This splitting is a signature of non-Gaussian fluctuations~\cite{Yan:2014nsa}.  
Non-Gaussianity is in fact expected in hydrodynamics because $v_2$ is proportional to the corresponding spatial anisotropy (denoted by $\varepsilon_2$) of the initial density profile~\cite{Gardim:2014tya}, and the fluctuations of $\varepsilon_2$ present generic non-Gaussian properties~\cite{Yan:2014afa,Gronqvist:2016hym}. 

In this article, we identify the main source of non-Gaussian 
fluctuations with the skewness of elliptic flow fluctuations in the reaction plane. 
We compute the skewness in event-by-event hydrodynamics (Sec.~\ref{s:hydro}) and compare it with the skewness of eccentricity fluctuations. 
We then show (Sec.~\ref{s:expansion}), by means of an expansion in powers of the fluctuations, that skewness is the leading contribution to the fine structure of higher-order cumulants. 
We compare experimental data with hydrodynamic calculations. 
In Sec.~\ref{s:gamma1}, we derive a general formula relating the standardized skewness to the first three cumulants,  $v_2\{2\}$, $v_2\{4\}$ and $v_2\{6\}$.  

\section{Skewness in event-by-event hydrodynamics}
\label{s:hydro}
In the flow picture~\cite{Luzum:2011mm}, particles are emitted independently in each collision with an azimuthal probability distribution, $P(\varphi)$, that fluctuates event to event.
We choose a coordinate frame where $\varphi=0$ is the direction of the reaction plane. 
Elliptic flow is defined as the second Fourier coefficient of
$P(\varphi)$, which has cosine and sine components: 
\begin{eqnarray}
\label{defvxy}
v_x&\equiv&\frac{1}{2\pi}\int_0^{2\pi}P(\varphi)\cos 2\varphi\, d\varphi , \cr
v_y&\equiv&\frac{1}{2\pi}\int_0^{2\pi}P(\varphi)\sin 2\varphi\, d\varphi .
\end{eqnarray}
Elliptic flow is a two-dimensional vector, ${\bf v}_2=v_x
{\bf e}_x+v_y{\bf e}_y$.
Using the standard terminology, we denote by $v_2$ the magnitude of ${\bf v}_2$, i.e. $v_2\equiv\sqrt{v_x^2+v_y^2}$.

Since the probability distribution, $P(\varphi)$, fluctuates event to event, the projections $v_x$ and $v_y$ are fluctuating quantities.
In hydrodynamics, these fluctuations result mainly from the fluctuations of the initial energy density profile and are due to the probabilistic nature of the positions of the nucleons within nuclei at the time of impact~\cite{Miller:2003kd,Alver:2006wh}.
${\bf v}_2$ is to a good approximation~\cite{Gardim:2014tya,Niemi:2012aj} proportional to the initial eccentricity ${\boldsymbol\varepsilon}_2=(\varepsilon_x,\varepsilon_y)$, which is defined by~\cite{Teaney:2010vd}: 
\begin{eqnarray}
\label{defepsxy}
\varepsilon_x&\equiv&-\frac{\int\rho(r,\phi)r^2\cos 2\phi\,
  rdrd\phi}{\int\rho(r,\phi)r^2\, rdrd\phi} , \cr 
\varepsilon_y&\equiv&-\frac{\int\rho(r,\phi)r^2\sin 2\phi\, rdrd\phi}{\int\rho(r,\phi)r^2\, rdrd\phi},
\end{eqnarray}
where $\rho(r,\phi)$ is the energy density deposited in the transverse plane shortly after the collision, in a centered polar coordinate
system. 

We model elliptic flow fluctuations by carrying out event-by-event
hydrodynamic calculations of Pb+Pb collisions at 2.76~TeV, with
initial conditions given by the Monte Carlo Glauber model~\cite{Alver:2008aq,Miller:2007ri,Broniowski:2007nz}. 
Our setup is the same as in Ref.~\cite{Noronha-Hostler:2015dbi}: The shear viscosity over entropy ratio is $\eta/s=0.08$~\cite{Policastro:2001yc} within the viscous relativistic hydrodynamical code {\footnotesize V-USPHYDRO} \cite{Noronha-Hostler:2013gga,Noronha-Hostler:2014dqa,Noronha-Hostler:2015coa}, which passes known analytical solutions \cite{Marrochio:2013wla},  
and $v_x$ and $v_y$ are calculated using Eq.~(\ref{defvxy}) at freeze-out~\cite{Teaney:2003kp} for pions in the transverse momentum range $0.2<p_t<3$~GeV/$c$.  

\begin{figure}[h]
\begin{center}
\includegraphics[width=\linewidth]{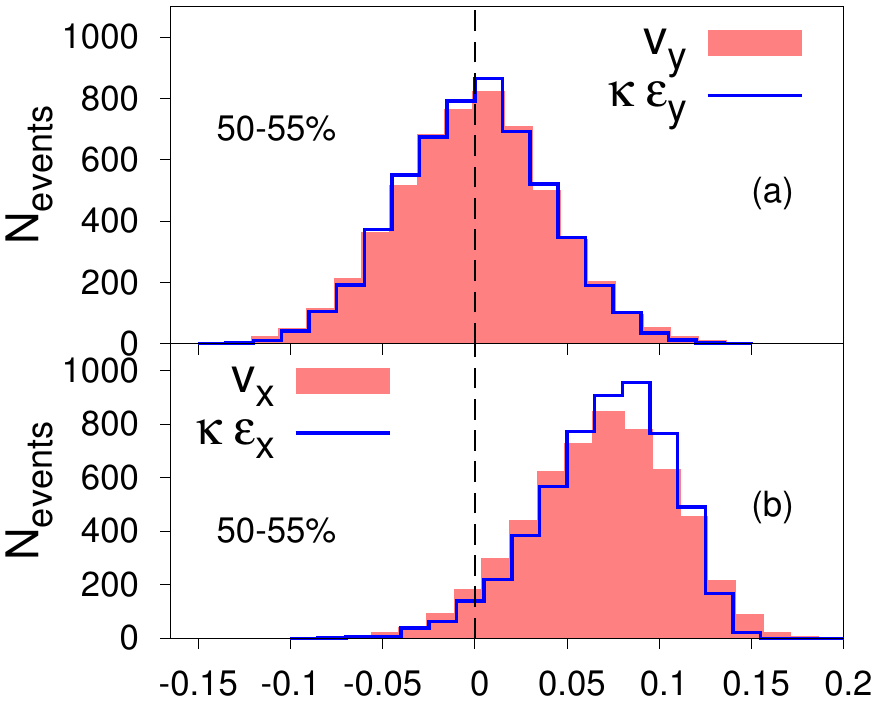} 
\end{center}
\caption{(Color online) 
\label{fig:histo}
Shaded areas: Histograms of the distribution of $v_y$ (a) and $v_x$ (b) for Pb+Pb collisions in the 50-55\% centrality range.  
5509 events were generated. 
Full lines: Histograms of the distributions of $\varepsilon_y$ (a) and $\varepsilon_x$ (b), rescaled  by a response coefficient $\kappa=0.21$. 
}
\end{figure} 
Figure~\ref{fig:histo} displays the histograms of the distributions of $v_y$ (a) and $v_x$ (b) in the 50-55\% centrality bin. 
We choose this rather peripheral centrality range as an illustration because elliptic flow is close to its maximum value~\cite{Aamodt:2010pa} and presents large fluctuations.  
Values of $v_x$ are positive for most events, corresponding to elliptic flow in the reaction plane~\cite{Ollitrault:1992bk}. 
We denote by $\bar v_2$ its mean value 
\begin{equation}
\label{order1}
\bar v_2 \equiv\langle v_x\rangle,
\end{equation}
where angular brackets denote an average over events in a centrality class. 
Note that $\bar v_2$ is smaller than the mean elliptic flow, $\langle v_2\rangle=\langle\sqrt{v_x^2+v_y^2}\rangle$. 
The distribution of $v_y$ is centered at $0$ because parity conservation and symmetry with respect to the reaction plane imply that the probability distribution of $(v_x,v_y)$ is symmetric under $v_y\to -v_y$. 
The magnitude of the fluctuations is characterized by the variances of $v_x$ and $v_y$:
\begin{eqnarray}
\label{order2}
\sigma_x^2&=&\langle (v_x-\bar v_2)^2\rangle=\langle v_x^2\rangle-\langle v_x\rangle^2 ,\cr
\sigma_y^2&=&\langle v_y^2\rangle.
\end{eqnarray}
For small fluctuations, the fluctuations of $v_x$ correspond to the fluctuations of the flow magnitude, while the fluctuations of $v_y$ correspond to the fluctuations of the flow angle.  
The so-called Bessel-Gaussian distribution~\cite{Voloshin:2007pc} of
$v_2$ is obtained by assuming that the distribution of ${\bf v}_2$ is
an isotropic two-dimensional Gaussian,  i.e., $\sigma_x=\sigma_y$.  
While this is typically a good approximation for central and mid-central collisions, it becomes worse as the centrality percentile increases. 
In particular, Fig.~\ref{fig:histo} shows that $\sigma_y$ is slightly
larger than $\sigma_x$, a general feature which can be traced back to the fluctuations of the initial eccentricity~\cite{Yan:2014afa}.  
The relative difference between $\sigma_y$ and $\sigma_x$ is in the fourth Fourier harmonic~\cite{Ollitrault:1997di} and, therefore, scales like $(\bar v_2)^2$. 

The distributions of $\varepsilon_x$ and $\varepsilon_y$ are also displayed in Fig.~\ref{fig:histo}, rescaled by a coefficient $\kappa$, so that the mean value of $\varepsilon_x$ matches that of the $v_x$ distribution. 
If ${\bf v}_2$ was linearly proportional to ${\boldsymbol\varepsilon}_2$, then the two distributions would be identical. 
The distribution of $v_x$ is somewhat broader than that of $\varepsilon_x$, mostly because of a cubic response term, which is expected to have a sizable contribution at large centrality~\cite{Noronha-Hostler:2015dbi}.  

One sees in Fig.~\ref{fig:histo} (b) that the distributions of $v_x$ and $\varepsilon_x$ are not symmetric with respect to their maximums: They present negative skew. 
The skewness of the distribution of $\varepsilon_x$ results from the condition $\varepsilon_x\le 1$, which acts as a right cutoff~\cite{Yan:2014afa}. 
Skewness is typically characterized by the third moment of the fluctuations. 
The symmetry $v_y\to -v_y$ allows for two non trivial moments to order 3: 
\begin{eqnarray}
\label{order3}
s_1&\equiv&\langle (v_x-\bar v_2)^3\rangle , \cr
s_2&\equiv&\langle (v_x-\bar v_2)v_y^2\rangle. 
\end{eqnarray}
The negative skew in Fig.~\ref{fig:histo} (b) corresponds to $s_1<0$. 
For dimensional reasons, a standardized skewness is usually employed, which is defined as
\begin{equation}
\label{gamma1}
\gamma_1\equiv\frac{s_1}{\sigma_x^3}. 
\end{equation}

\begin{figure}[h]
\begin{center}
\includegraphics[width=\linewidth]{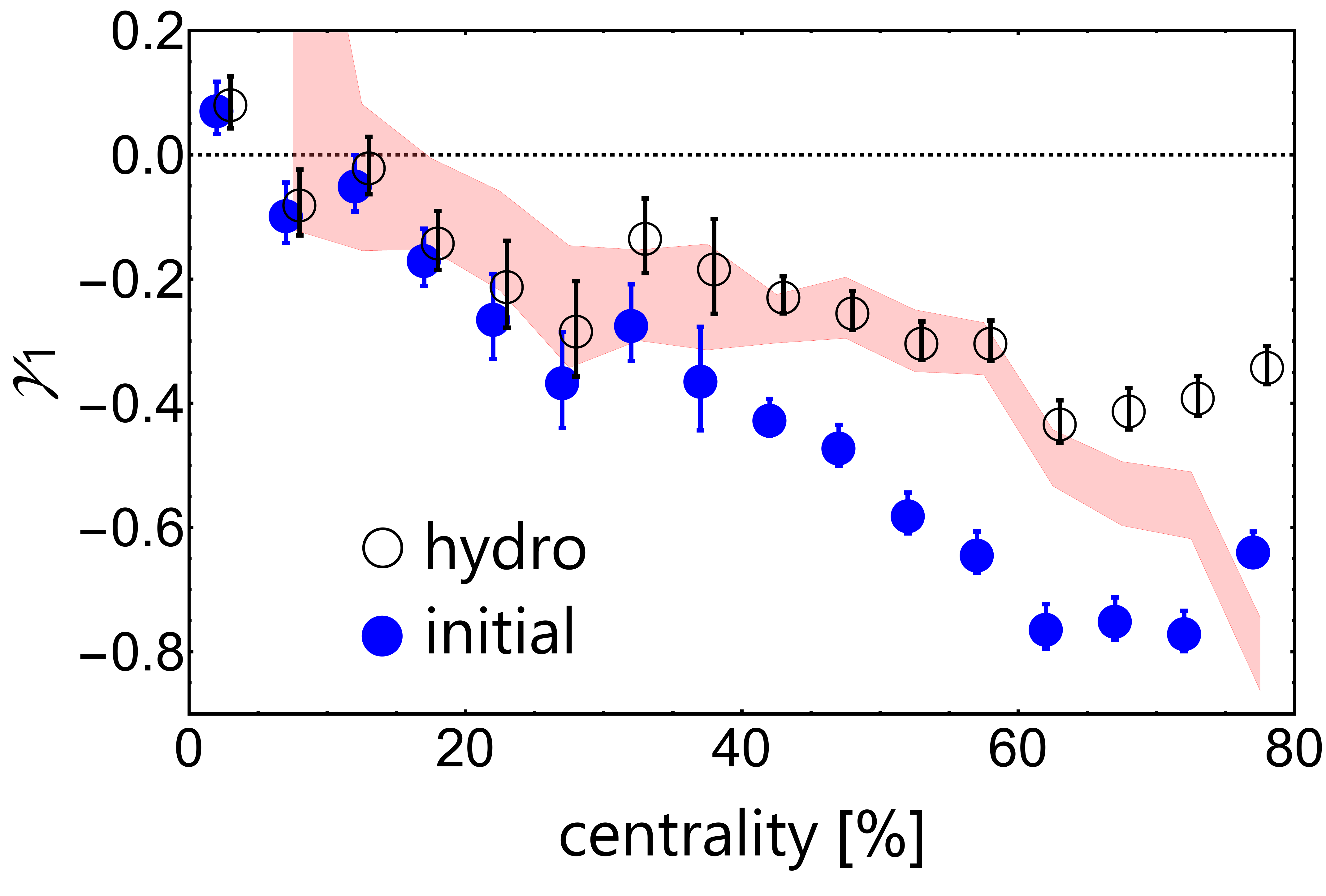} 
\end{center}
\caption{(Color online) 
\label{fig:truegamma}
Standardized skewness of elliptic flow fluctuations (open circles) and of initial eccentricity fluctuations (full circles) from hydrodynamic calculations, as a function of centrality percentile, for Pb+Pb collisions at 2.76~TeV. 
Symbols have been slightly shifted horizontally for the sake of readability.  
The shaded band displays the value of $\gamma_1$ estimated from the cumulants of $v_2$, as defined by Eq.~(\ref{gamma1exp}). 
}
\end{figure} 

Figure~\ref{fig:truegamma} displays the standardized skewness, $\gamma_1$, calculated in hydrodynamics as a function of the collision centrality.
It is negative above $15\%$ centrality and its absolute magnitude increases as a function of centrality percentile. 
This increase results from two effects: First, $\gamma_1$ vanishes by symmetry for central collisions and is typically proportional to $\bar v_2$; second, it is a first-order correction to the central limit and is, therefore, inversely proportional to the square root of the system size~\cite{Gronqvist:2016hym}.
Figure~\ref{fig:truegamma} also displays the standardized skewness of the $\varepsilon_x$ fluctuations, which, as we pointed out before, would be identical to that of the $v_x$ fluctuations if ${\bf v}_2$ were exactly linearly proportional to ${\boldsymbol\varepsilon}_2$. 
We observe that the standardized skewness calculated from ${\bf v}_2$ becomes smaller in absolute value than the initial skewness calculated from ${\boldsymbol\varepsilon}_2$ as the centrality percentile increases.
Hence, the hydrodynamical evolution washes out part of the initial skewness. 
This effect, which is clearly seen in the histogram of Fig.~\ref{fig:histo}, is mostly due to the cubic response of the system, which increases $\sigma_x$~\cite{Noronha-Hostler:2015dbi}. 

Equations~(\ref{order1})--(\ref{order3}) are the first-order terms in a cumulant expansion of the flow fluctuations. 
The formalism of generating functions provides a compact formulation for the cumulant expansion. 
The Fourier-Laplace transform of the distribution of ${\bf v_2}$ is 
$\langle e^{{\bf k}\cdot{\bf v}_2}\rangle$, where ${\bf k}\equiv k_x
{\bf e}_x+k_y{\bf e}_y$ is a two-dimensional vector. 
The generating function of the cumulants is its logarithm, $\ln\langle
e^{{\bf k}\cdot{\bf v}_2}\rangle$. 
By expanding it up to order 3 in ${\bf k}$, one obtains 
\begin{equation}
\label{cumulantframe1}
\ln\langle  e^{{\bf k}\cdot{\bf v}_2}\rangle=
k_x\bar v_2+\frac{k_x^2}{2}\sigma_x^2+\frac{k_y^2}{2}\sigma_y^2
+\frac{k_x^3}{6}s_1+\frac{k_xk_y^2}{2}s_2 ,
\end{equation}
where $\bar v_2$, $\sigma_x$, $\sigma_y$, $s_1$, and $s_2$ are given by
Eqs.~(\ref{order1})--(\ref{order3}). 

\section{The fine structure of higher-order cumulants}
\label{s:expansion}

The direction of the reaction plane is not known experimentally.
Therefore, the skewness of the $v_x$ fluctuations defined in Eq.~(\ref{gamma1}) cannot be measured directly. 
More specifically, there is no simple way of extracting it from the probability distribution of the flow magnitude, $v_2$~\cite{Aad:2013xma}. 
In this section, we show how one can relate the skewness to quantities
which are measured experimentally, specifically, the cumulants of the
distribution of $v_2$.  

Experimental observables are measured in the laboratory frame where the orientation of the reaction plane has a
flat distribution.
The cumulants of the distribution of $v_2$, as measured in
experiments~\cite{Aamodt:2010pa,Aad:2014vba,Adler:2002pu,Alt:2003ab,Chatrchyan:2012ta}, 
are defined in this frame~\cite{Borghini:2000sa,Borghini:2001vi}. 
Their generating function is given by the left-hand side of Eq.~(\ref{cumulantframe1}), with the only difference that one averages over the orientation of the reaction plane before taking the logarithm: 
One exponentiates Eq.~(\ref{cumulantframe1}), substitutes $k_x=k\cos\varphi$ and $k_y=k\sin\varphi$, averages over $\varphi$, and finally takes the logarithm:
\begin{equation}
\label{cumulantframe2}
\ln G(k)\equiv\ln\left(\int_0^{2\pi}\frac{d\varphi}{2\pi}\langle
\textup{e}^{{\bf k}\cdot{\bf v}_2}\rangle\right).
\end{equation}
The $2n$-th order cumulant, $v_2\{2n\}$, is eventually given by the
$2n$-th order term of the Taylor expansion of $\ln G({k})$ computed at
$k=0$\footnote{In the Taylor expansion we consider only terms of order
  $2n$ because $I_0(k)$ is even.}.
More specifically~\cite{Borghini:2001vi}:
\begin{equation}
\label{defv2n}
\left.\frac{d^{2n}}{dk^{2n}}\ln I_0(kv_2\{2n\})\right|_{k=0}\equiv
\left.\frac{d^{2n}}{dk^{2n}}\ln G(k)\right|_{k=0}.
\end{equation}
In the simple case of Bessel-Gaussian fluctuations, $s_1=s_2=0$ and $\sigma_y=\sigma_x$.
Inserting Eq.~(\ref{cumulantframe1}) into Eq.~(\ref{cumulantframe2}), one obtains
\begin{equation}
\ln G(k)=\ln I_0(k\bar v_2)+\frac{k^2\sigma_x^2}{2},
\end{equation}
and Eq.~(\ref{defv2n}) yields 
\begin{eqnarray}
\label{gaussian}
v_2\{2\}&=&\sqrt{(\bar v_2)^2+2\sigma_x^2} , \cr
v_2\{4\}=v_2\{6\}=\cdots&=&\bar v_2. 
\end{eqnarray}
Therefore, the cumulants of order $n\ge 4$ are identical to the mean
elliptic flow in the reaction plane~\cite{Voloshin:2007pc}. 

\begin{figure}[h]
\begin{center}
\includegraphics[width=\linewidth]{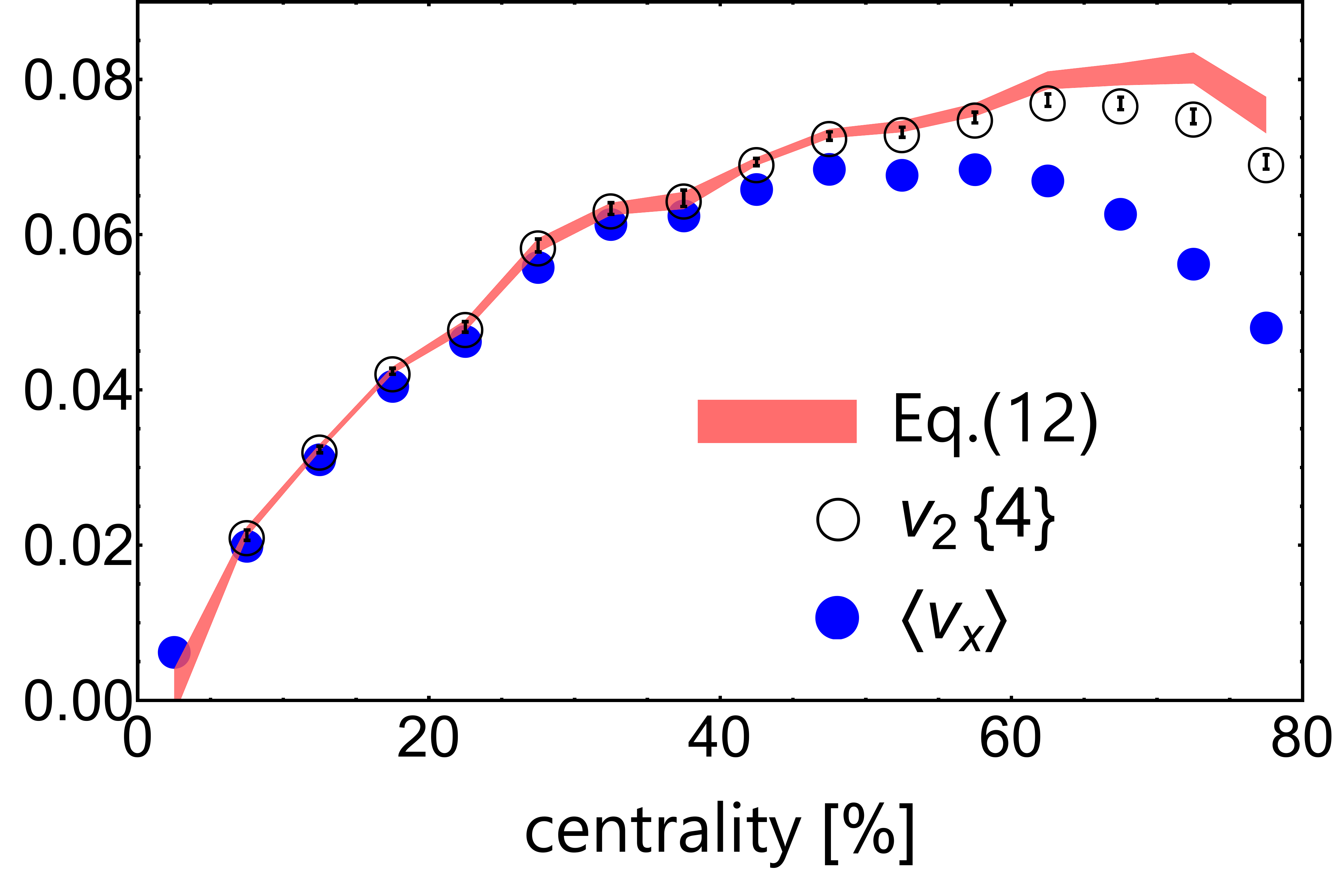} 
\end{center}
\caption{(Color online) 
Open symbols: $v_2\{4\}$ vs centrality in event-by-event
hydrodynamics. 
Full symbols: mean elliptic flow in the reaction plane $\langle
v_x\rangle=\bar v_2$. 
Shaded band: right-hand side of Eq.~(\ref{asymskew}) for $v_2\{4\}$,
corresponding to the leading non-Gaussian corrections. 
\label{fig:meanv2x}
}
\end{figure}

In event-by-event hydrodynamics, the direction of the reaction plane
is known and one can compute both
$v_2\{4\}$~\cite{Qiu:2011iv,Hirano:2012kj,Bozek:2013uha,Niemi:2015qia}
and $\bar v_2$. 
Figure~\ref{fig:meanv2x} shows their dependence on the
centrality percentile. They are compatible up to $40\%$
centrality. For peripheral collisions, $v_2\{4\}$ becomes
significantly larger than $\bar v_2$, which means that the
Bessel-Gaussian ansatz fails~\cite{Qiu:2011iv}.  
This failure can be attributed either to the asymmetry of the
fluctuations, $\sigma_x\not=\sigma_y$, or to non-Gaussian
fluctuations. 
Both these features are expected in hydrodynamics, as shown in Sec.~\ref{s:hydro}. 
Expanding the generating function in powers of the fluctuations and keeping only the leading order terms in $\sigma_y^2-\sigma_x^2$, $s_1$ and $s_2$, we obtain:
\begin{eqnarray}
\label{asymskew}
v_2\{2\}&=&\sqrt{(\bar v_2)^2+\sigma_x^2+\sigma_y^2} ,\cr
v_2\{4\}&\simeq&\bar v_2+\frac{\sigma_y^2-\sigma_x^2}{2\bar
  v_2}-\frac{s_1+s_2}{(\bar v_2)^2} ,\cr
v_2\{6\}&\simeq&\bar v_2+\frac{\sigma_y^2-\sigma_x^2}{2\bar
  v_2}-\frac{\frac{2}{3}s_1+s_2}{(\bar v_2)^2} ,\cr
v_2\{8\}&\simeq&\bar v_2+\frac{\sigma_y^2-\sigma_x^2}{2\bar
  v_2}-\frac{\frac{7}{11}s_1+s_2}{(\bar v_2)^2} ,
\end{eqnarray}
When these corrections are added, higher-order cumulants are no longer
equal to $\bar v_2$. The shaded band in Fig.~\ref{fig:meanv2x}
corresponds to the right-hand side of the second line of Eq.~(\ref{asymskew}), where all
terms are calculated in hydrodynamics. Agreement with the left-hand
side is excellent for all centralities. The term proportional to the 
asymmetry of the fluctuations, $\sigma_y^2-\sigma_x^2$, turns out to
be negligible: The leading 
correction is the term proportional to $s_1+s_2$, due to the
non-Gaussianity of the fluctuations. 

\begin{figure}[h]
\begin{center}
\includegraphics[width=.97\linewidth]{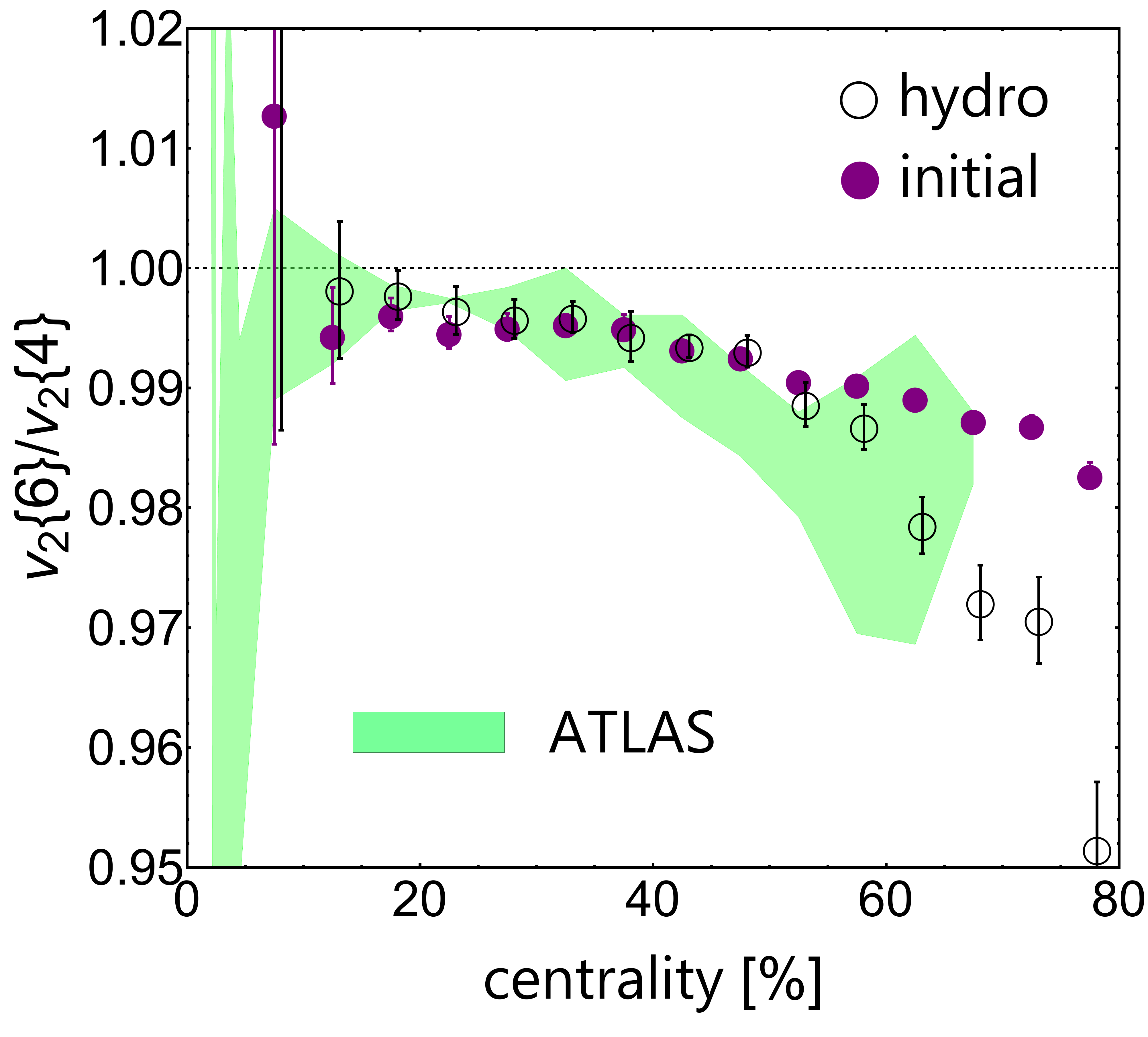} 
\end{center}
\caption{(Color online) 
\label{fig:cumulants}
Shaded band: ATLAS data for $v_2\{6\}/v_2\{4\}$ versus
centrality~\cite{Aad:2013xma}. 
Error bars take into account the strong correlation
between $v_2\{6\}$ and $v_2\{4\}$~\cite{Borghini:2001vi}. 
Open symbols: hydrodynamic calculations. 
Full symbols: $\varepsilon_2\{6\}/\varepsilon_2\{4\}$. 
}
\end{figure} 

Non-Gaussian fluctuations not only increase the value of $v_2\{4\}$: 
They also induce a splitting between $v_2\{4\}$, $v_2\{6\}$ and  
$v_2\{8\}$.
Subtracting the second and third line of Eq.~(\ref{asymskew}), one obtains: 
\begin{equation}
\label{effectofskewness}
v_2\{4\}-v_2\{6\}=-\frac{s_1}{3(\bar v_2)^2}.
\end{equation}
The splitting is solely due to the coefficient $s_1$, corresponding to 
the skewness of elliptic flow fluctuations in the reaction 
plane.\footnote{When higher-order corrections are taken into account,
  the asymmetry between $\sigma_y$ and $\sigma_x$ also   produces a
  splitting between $v_2\{4\}$ and $v_2\{6\}$, of 
  order $(\sigma_y^2-\sigma_x^2)^3$; the corresponding contribution 
  is much smaller than that of $s_1$ and $s_2$ and has opposite sign.} 
Figure~\ref{fig:cumulants} displays ATLAS data for
$v_2\{6\}/v_2\{4\}$ versus centrality for Pb+Pb collisions at
2.76~TeV. We use the data from Fig.~9b of 
  Ref.~\cite{Aad:2014vba}, inferred from the event-by-event
  distribution of $v_2$~\cite{Aad:2013xma}, which have smaller error bars than
  the direct cumulant measurements.
$v_2\{4\}$ and $v_2\{6\}$ are very close to one another, but 
  one observes a fine structure, at the percent 
  level, for most centralities: $v_2\{4\}$ is larger than $v_2\{6\}$.
  This, according to Eq.~(\ref{effectofskewness}), implies $s_1<0$, 
in line with our expectation from the hydrodynamic calculations presented in Sec.~\ref{s:hydro}. 
We carry out a more quantitative comparison by numerical calculations of
$v_2\{6\}/v_2\{4\}$ in hydrodynamics. The result is displayed in Fig.~\ref{fig:cumulants} (open symbols). 
It is compatible with experimental data within error bars. 
Precise figures depend on the model of initial conditions, but Fig.~\ref{fig:cumulants} shows that hydrodynamics naturally captures the skewness of the $v_2$ fluctuations, hence the splitting between $v_2\{4\}$ and $v_2\{6\}$. 

In our hydrodynamic calculation, the ratio 
$v_2\{6\}/v_2\{4\}$ coincides with the corresponding ratio for initial eccentricities, $\varepsilon_2\{6\}/\varepsilon_2\{4\}$, up to 60\% centrality.\footnote{We do not have a simple explanation for the difference above 60\%
centrality. It is a nonlinear hydrodynamic effect. However, we have
checked that it is not captured by the cubic response alone.} 
We stress that this was not \textit{a priori} expected because the cubic response
breaks simple proportionality and decreases the skewness of the
distribution of $v_2$ compared to that of $\varepsilon_2$. 
While the cubic response has an important effect on the ratio 
$v_2\{4\}/v_2\{2\}$~\cite{Noronha-Hostler:2015dbi}, 
it does not seem to affect the ratio $v_2\{6\}/v_2\{4\}$, which directly reflects the ratio $\varepsilon_2\{6\}/\varepsilon_2\{4\}$ provided by the model of initial conditions. 

Equation~(\ref{asymskew}) also gives the following universal prediction for the small splitting between $v_2\{6\}$ and $v_2\{8\}$:\footnote{Results similar to Eqs.~(\ref{effectofskewness}) and
(\ref{v8}) have been obtained \cite{Jia:2014pza} by 
studying the distribution of $v_2$ in the limit of small fluctuations.}
\begin{equation}
\label{v8}
v_2\{6\}-v_2\{8\}=\frac{1}{11}(v_2\{4\}-v_2\{6\}).
\end{equation}
The number of events in our hydrodynamic calculation is too small to
test this relation. 
However, the same relation can be written for the cumulants of the 
initial eccentricity, $\varepsilon_2$. It is obtained by replacing  
$v_2$ with $\varepsilon_2$ everywhere in the derivation, and thus does
not involve any relation between $\varepsilon_2$ and $v_2$. 
We have tested Eq.~(\ref{v8}) for the fluctuations of $\varepsilon_2$ within a 
Monte Carlo Glauber model, which allows for much higher statistics than 
full hydrodynamic calculations. 
We find that Eq.~(\ref{v8}) is approximately satisfied for central collisions, but that the left-hand side becomes larger than the right-hand side as the centrality percentile increases. 
This means that the expansion leading to Eq.~(\ref{v8}) is unable to
capture accurately the splitting between $\varepsilon_2\{6\}$ and
$\varepsilon_2\{8\}$, and consequently the
splitting between $v_2\{6\}$ and $v_2\{8\}$. 

\section{Measuring the skewness with cumulants}
\label{s:gamma1}
In this section we explain how to estimate the standardized skewness, $\gamma_1$, defined in Eq.~(\ref{gamma1}), from $v_2\{2\}$, $v_2\{4\}$, and $v_2\{6\}$.  
We estimate $s_1$ using Eq.~(\ref{effectofskewness}). 
Since this result is derived from a perturbative expansion to first
order in $s_1$, we estimate also $\gamma_1$ to first order.  
By doing so, we neglect small non-Gaussian contributions to $\bar v_2$
and $\sigma_x$:
We use the Gaussian approximation, Eq.~(\ref{gaussian}), which gives
\begin{eqnarray}
\label{gaussianapprox}
v_2\{4\}&=&\bar v_2 , \cr
v_2\{2\}^2-v_2\{4\}^2&=&2\sigma_x^2.
\end{eqnarray}
Using Eqs.~(\ref{effectofskewness}) and (\ref{gaussianapprox}), we  obtain the following estimate of $\gamma_1$, which we denote by $\gamma_1^{\rm expt}$:  
\begin{equation}
\label{gamma1exp}
\gamma_{1}^{\rm expt}\equiv -6\sqrt{2}\, v_2\{4\}^2\frac{v_2\{4\}-v_2\{6\}}{(v_2\{2\}^2-v_2\{4\}^2)^{3/2}}.
\end{equation}

\begin{figure}[h]
\begin{center}
\includegraphics[width=\linewidth]{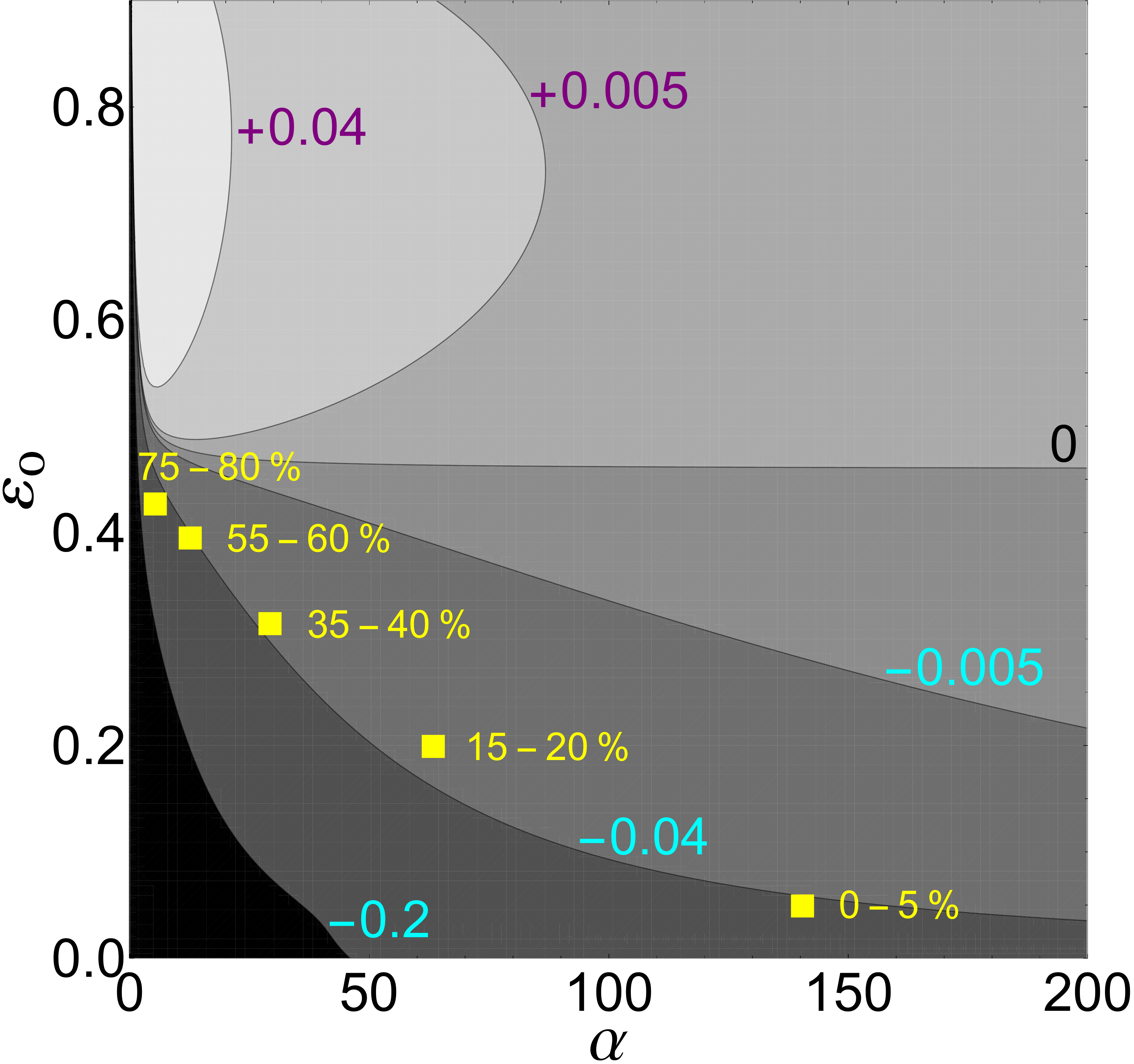} 
\end{center}
\caption{(Color online) 
\label{fig:ellipticpower}
Contour plot of the difference $\gamma_1^{\rm expt}~-~\gamma_1$, with $\gamma_1^{\rm expt}$ defined in Eq.~(\ref{gamma1exp}) and $\gamma_1$ defined in Eq.~(\ref{gamma1}), computed by means of the elliptic-power distribution~\cite{Yan:2014afa}, in the $(\alpha,\varepsilon_0)$ parameter plane. 
Squares correspond to the values of $\alpha$ and $\varepsilon_0$ 
extracted from Monte Carlo Glauber~\cite{Alver:2008aq} simulations of Pb+Pb 
collisions at 2.76~TeV, which are fitted to the elliptic-power
distribution.  
}
\end{figure} 
We check the accuracy of $\gamma_1^{\rm expt}$  as an estimate of $\gamma_1$ using two different methods. 
The first method is to compute both $\gamma_1$ and $\gamma_1^{\textup{expt}}$ in event-by-event hydrodynamics. 
$\gamma_1^{\rm expt}$ is shown as a shaded band in
Fig.~\ref{fig:truegamma}. 
It is in good agreement with $\gamma_1$ up to 60\% centrality. 
Above 60\% centrality, the approximation $v_2\{4\}\simeq \bar v_2$
breaks down, as shown by Fig.~\ref{fig:meanv2x}. 
Statistical errors in our hydrodynamic calculation are significant due to the limited amount of events in each centrality bin. 
Therefore we employ a second method. 
Since Eq.~(\ref{gamma1exp}) can be derived as well for the
skewness of the distribution of $\varepsilon_2$, we test the validity
of this relation using the elliptic-power distribution \cite{Yan:2014afa}, which is a simple analytical model for the distribution of $(\varepsilon_x, \varepsilon_y)$.
The elliptic-power distribution has two parameters: $\varepsilon_0$, which approximately gives the mean eccentricity in the reaction plane, $\varepsilon_0\simeq \langle\varepsilon_x\rangle$, and $\alpha$, which is proportional to the number of participants.  
We evaluate both $\gamma_1$ and  $\gamma_1^{\rm expt}$ as a function of
$\varepsilon_0$ and $\alpha$. 
Fluctuations scale like $1/\sqrt{\alpha}$, therefore, the assumption of small fluctuations made in deriving Eq.~(\ref{asymskew}) holds for $\alpha\gg 1$. 
One also expects approximations to break down in the limit 
$\varepsilon_0\to 0$ (corresponding to the limiting case of the power
distribution~\cite{Yan:2013laa}) where $\gamma_1$ vanishes by symmetry
while $\gamma_1^{\rm expt}$ does not.
Figure~\ref{fig:ellipticpower} indeed shows that the difference between the
estimated skewness and the true skewness is large only when both
$\alpha$ and $\varepsilon_0$ are small. 
In order to estimate the range of $\alpha$ and $\varepsilon_0$
applicable to Pb+Pb collisions, we perform Monte Carlo
Glauber~\cite{Alver:2008aq} simulations and fit the resulting distribution of
$\varepsilon_2$ to the elliptic-power distribution, for different centrality
windows. The values of $\alpha$ and $\varepsilon_0$ extracted from the fits are shown as squares in
Fig.~\ref{fig:ellipticpower}. 
Based on this figure, and since in hydrodynamics the skewness of $v_2$ is
comparable to that of $\varepsilon_2$, we expect the difference
$|\gamma_1^{\rm expt}-\gamma_1|$ to be a few $10^{-2}$ for
Pb+Pb collisions, much smaller in absolute value than the value of $\gamma_1$ in Fig.~\ref{fig:truegamma}. 
Therefore, Eq.~(\ref{gamma1exp}) should provide a reasonable estimate of the standardized skewness also from experimental data.

\begin{figure}[h]
\begin{center}
\includegraphics[width=\linewidth]{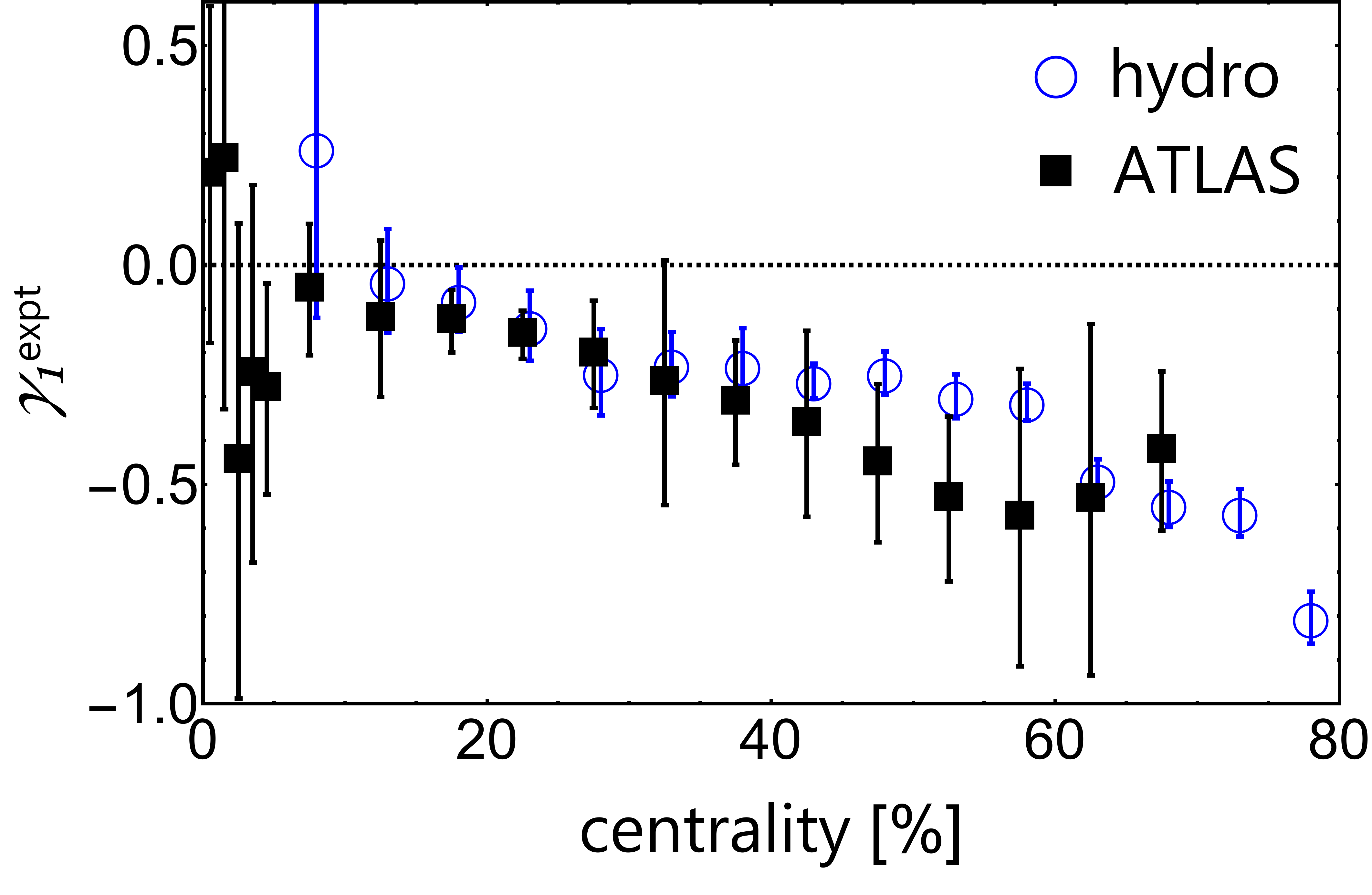} 
\end{center}
\caption{(Color online) 
Standardized skewness of $v_2$ fluctuations, as defined in Eq.~(\ref{gamma1exp}), as a function of centrality. 
Squares: ATLAS data. 
Circles: hydrodynamic calculations, corresponding to the dark shaded band
in Fig.~\ref{fig:truegamma}.  
Symbols have been slightly shifted horizontally for the sake of readability. 
\label{fig:expskewness}
}
\end{figure} 
Figure~\ref{fig:expskewness} displays the skewness extracted from
ATLAS data \cite{Aad:2013xma} using Eq.~(\ref{gamma1exp}). 
The standardized skewness is moderate but not small, and reaches $-0.5$ in peripheral collisions, although with large error bars. 
Errors have been estimated by adding statistical and systematic
errors in quadrature, and assuming that the errors on $v_2\{2\}$,
$v_2\{4\}$, and $v_2\{6\}/v_2\{4\}$ are uncorrelated. 
Since errors on $v_2\{2\}$ and $v_2\{4\}$ are usually correlated, the
errors on ATLAS data in Fig.~\ref{fig:expskewness} are probably overestimated.
Hydrodynamic calculations are compatible with experimental data in the full range of centrality.

\section{Conclusions}
We have shown that the small splitting of  higher-order cumulants of the elliptic flow from mid-central up to peripheral ultrarelativistic nucleus-nucleus collisions is mostly due to the skewness of the fluctuations of the elliptic flow in the reaction plane, $v_x$. 
We emphasize that this is a general result which does not depend on
any particular model. 
Negative skewness is observed in Pb+Pb data, and is naturally
explained in hydrodynamics: it follows from the fact that $v_2$  
is approximately proportional to the initial eccentricity, and that
the eccentricity in the reaction plane is bounded by unity. 
The splitting between $v_2\{4\}$ and $v_2\{6\}$ 
thus provides additional evidence of the collective origin of elliptic flow.
We have computed the ratio $v_2\{6\}/v_2\{4\}$ in event-by-event
viscous hydrodynamics and we have shown that it is very close to the ratio $\varepsilon_2\{6\}/\varepsilon_2\{4\}$ between the cumulants of the initial eccentricity. 
Thus, this observable constrains the early dynamics of the quark-gluon plasma~\cite{Hirano:2005xf,Bhalerao:2011yg,Schenke:2012wb,Albacete:2013tpa,Renk:2014jja}.

\section*{Acknowledgements}
This work is 
supported by the European Research Council under the
Advanced Investigator Grant ERC-AD-267258. 
JNH acknowledges the use of the Maxwell Cluster and the advanced support from the Center of Advanced Computing and Data Systems at the University of Houston to carry out the research presented here. JNH~was supported by the National Science Foundation under grant no.~PHY-1513864
and underneath FAPESP grant: 2016/03274-2.

\end{document}